\documentclass[aps,prd,preprint,superscriptaddress]{revtex4-1}
\usepackage[mathscr]{euscript}
\usepackage{float,epsfig}
\usepackage[normalem]{ulem}
\usepackage{bm}
\usepackage{graphicx}
\usepackage{subfigure}
\usepackage{appendix}
\usepackage{amsmath,amssymb,amsthm}
\usepackage[colorlinks=true,linkcolor=blue,urlcolor=blue]{hyperref}
\newcommand{\bea}{\begin{eqnarray}}
\newcommand{\eea}{\end{eqnarray}}
\newcommand{\beq}{\begin{equation}}
\newcommand{\eeq}{\end{equation}}

\def\/{\over}

\begin{document}


\title{Induced quantum gravitational interaction between two objects with permanent quadrupoles in external gravitational fields }

\author{Yongshun Hu}
\affiliation{Department of Physics and Synergetic Innovation Center for Quantum Effects and Applications, Hunan Normal University, Changsha, Hunan 410081, China}
\author{Jiawei Hu}
\email[Corresponding author. ]{jwhu@hunnu.edu.cn}
\affiliation{Department of Physics and Synergetic Innovation Center for Quantum Effects and Applications, Hunan Normal University, Changsha, Hunan 410081, China}
\author{Hongwei Yu}
\email[Corresponding author. ]{hwyu@hunnu.edu.cn}
\affiliation{Department of Physics and Synergetic Innovation Center for Quantum Effects and Applications, Hunan Normal University, Changsha, Hunan 410081, China}

\begin{abstract}

We investigate, in the framework of linearized quantum gravity, the induced quantum gravitational interaction between two ground-state objects with permanent quadrupole moments, which are subjected to an external gravitational radiation field. 
Compared with the nonpolar case, there exists an additional term in the leading-order field-induced interobject interaction between two polar objects. 
This term arises when a real graviton is scattered by the same object center with coupling between the two objects occurring via the exchange of a single virtual graviton,
and the interaction is thus relevant to the number density of gravitons, the frequency and polarization of the external gravitational field, as well as the permanent quadrupoles of the objects. 
Due to the existence of such an additional term, the field-induced quantum gravitational interaction between two polar objects can be significantly different from that between nonpolar ones when the interobject distance is much smaller than the wavelength of the  external gravitational radiation field.
Although we model the external gravitational radiation as a quantized monochromatic gravitational wave with a certain wave vector and polarization for simplicity, it is possible to generalize to more realistic situations, such as the case of a stochastic background of gravitational radiation.

\end{abstract}


\maketitle

\section{Introduction}
\label{sec_in}
\setcounter{equation}{0}
Gravitational waves, which are ripples of spacetime predicted naturally by general relativity~\cite{Einstein1916}, may be regarded as a number of gravitons propagating through the universe when gravity is quantized. In this respect, a keen interest in the possible quantum nature or quantum properties of gravitational waves arises. Unfortunately, a full theory of quantum gravity is elusive at present. Nonetheless, one can still study  low energy quantum gravitational effects,  since, at low energy scales, general relativity can be treated as an effective field theory. For example, the quantum correction to the Newtonian potential between two point masses has been obtained by summing one-loop Feynman diagrams with off-shell gravitons~\cite{Donoghue1994prl,Donoghue1994prd,Hamber1995,Kirilin2002,Holstein2003,Holstein2005}. Also, a framework of linearized quantum gravity has been established in the study of the quantum light-cone fluctuations~\cite{Ford1995,Ford1996,yu1999,yu2000,yu2009}, the basic idea of which is to quantize the linearized perturbation propagating on a flat background spacetime in the canonical approach.

Recently, based on linearized quantum gravity, the quantum  gravitational  vacuum fluctuation induced 
interaction between two nonpointlike objects in their ground states as well as that between a nonpointlike object and a gravitational boundary have been studied in Refs. \cite{Ford2016,Wu2016,Wu2017,Holstein2017,Hu2017,yu2018}.
Later on, in analogy to the cases in electrodynamics \cite{CP,PT,Salam,Thirunamachandran1980,Milonni1992,Milonni1996,Salam2006,Buhmann2019}, the behavior of the interobject vacuum-fluctuation-induced quadrupole-quadrupole interactions is found to be relative to the quantum states of the objects \cite{yongs2020epjc}. 
That is, for instance, the quantum gravitational quadrupole-quadrupole interaction  behaves as $r^{-10}$ and $r^{-11}$ in the near and far regimes respectively when the two objects are in their ground states \cite{Ford2016,Wu2016,Wu2017,Holstein2017}, while it behaves as $r^{-5}$ and $r^{-1}$ in the near and far regimes respectively when the two objects are in a symmetric/antisymmetric  entangled state \cite{yongs2020epjc}.
Moreover, in the presence of an external gravitational radiation field, the interobject quadrupole-quadrupole interaction between two ground-state objects is also found to be significantly different from that of the vacuum case. It has been demonstrated  in Ref. \cite{yongs2020prd} that the external gravitational field-induced interobject interaction behaves as $r^{-5}$ in the near regime and oscillates  with a decreasing amplitude proportional to $r^{-1}$ in the far regime, and whether the interaction is attractive or repulsive depends on the propagation direction, polarization and frequency of the external gravitational field.

The  quantum  gravitational interaction discussed above is  obtained under the assumption that the  objects  have no permanent quadrupole moments. Since polar objects exist extensively in the universe, a natural question arises as to what the  quantum gravitational interaction will be in the presence of external gravitational radiation if objects with permanent quadrupole moments are concerned.
Similar examples in quantum electrodynamics show that, compared to the nonpolar case,
in which the  field-induced interatomic or intermolecular  interaction arises when a real photon is scattered by the two atomic or molecular centers with coupling between the bodies occurring via the exchange of a single virtual photon, there is an additional term in  the field-induced interatomic or intermolecular interaction between atoms or molecules with permanent dipole moments 
\cite{Andrews2000,Andrews2005,Salam2007}, which arises when a real photon is scattered by the same atomic or molecular center. 
Likewise, we expect that  in the gravitational case, the quantum interobject gravitational interaction between objects with permanent quadrupole moments should also be discriminative from that between nonpolar objects.

In this paper, we study the induced quantum gravitational interaction between two ground-state objects with permanent quadrupole moments in  a weak external gravitational radiation field, in the framework of linearized quantum gravity. 
First, we give a derivation of the induced interaction energy shift between two ground-state objects with  permanent quadrupole moments  based on the fourth-order perturbation theory. Then,  we compare 
the result with that in the nonpolar case given in Ref. \cite{yongs2020prd}.
For simplicity, we model the external gravitational radiation as a quantized monochromatic gravitational wave with a certain wave vector and polarization. However, it is possible to generalize to more realistic situations. As an example, a generalization to the case of a stochastic background of gravitational radiation is discussed.
Throughout the paper, the Latin indices run from $1$ to $3$, the Greek indices run from $0$ to $3$, and the Einstein summation convention is assumed.

\section{Basic equations}
\label{sec_ge}
We consider two nonpointlike objects A and B coupled with the fluctuating gravitational fields in vacuum, which are endowed  with permanent quadrupole moments, and are subjected  to an externally applied  weak  gravitational radiation field.
The two objects (A and B) are modeled as multilevel systems with  the ground and excited energy eigenvalues being $E_0$ and $E_{s} ~( s=1,2,3,...)$, and the corresponding eigenstates being $|e_0\rangle$ and $|e_s\rangle$, respectively. The total Hamiltonian of the system considered is
\beq \label{Hamiltonian}
H=H_S+H_F+H_R+H_I,
\eeq
where $H_{S}$ is the Hamiltonian of the two objects A and B, $H_{F}$  the Hamiltonian of the fluctuating gravitational fields, $H_R$  the Hamiltonian of the external gravitational radiation field, and $H_I$  the interaction Hamiltonian between the objects and the gravitational fields, which takes the form
\beq\label{HI}
H_I=-\frac{1}{2}Q^{A}_{ij}[\epsilon_{ij}(\vec x_A)+E_{ij}(\vec x_A)]-\frac{1}{2}Q^{B}_{ij}[\epsilon_{ij}(\vec x_B)+E_{ij}(\vec x_{B})].
\eeq
Here $Q^{\xi}_{ij}$ is the quadrupole moment operator of object $\xi$ ($\xi=$ A, B), $\epsilon_{ij}(\vec x)$  the gravitoelectric tensor of the external gravitational radiation field, and $E_{ij}(\vec x)$ the gravitoelectric tensor characterizing the fluctuating gravitational fields in vacuum, which is defined by an analogy between the linearized Einstein field equations and the Maxwell's equations as $E_{ij}=-c^2 C_{0i0j}$ ~\cite{Campbell1971,Campbell1976,Maartens1998,Matte1953,Ramos2010,Szekeres1971,Ruggiero2002}, where $C_{\alpha\beta\mu\nu}$ is the Weyl tensor and $c$ the speed of light.
In the absence of external gravitational radiation, the metric tensor of the spacetime $g_{\mu\nu}$ can be expressed as $g_{\mu\nu}=\eta_{\mu\nu}+h_{\mu\nu}$, with $\eta_{\mu\nu}$ being the flat spacetime metric and $h_{\mu\nu}$ the fluctuating gravitational fields. Then, in the transverse traceless gauge, the gravitoelectric tensor $E_{ij}$ is found to be $E_{ij}=\frac{1}{2}\ddot h_{ij}$, which  can be quantized in the canonical approach  as
\beq\label{Eij}
E_{ij}
=-\sum_{\vec p,\lambda} \sqrt{\frac{\hbar G\omega^3}{c^2(2\pi)^2}} \left[a_{\lambda}(\vec p) e^{(\lambda)}_{ij} e^{i(\vec p\cdot \vec x-\omega t)}+\text{H.c.}\right],
\eeq
where 
$\hbar$ is the reduced Planck constant, and $G$ the Newtonian gravitational constant. Here $a_{\lambda}(\vec p)$ is the annihilation operator of the fluctuating gravitational fields, $\lambda$ labels the polarization states, $e^{(\lambda)}_{ij}$ are polarization tensors, $\omega=c|\vec p|=c(p_{x}^2+p_{y}^2+p_{z}^2)^{1/2}$, and H.c. denotes the Hermitian conjugate. 
Assume that the weak external gravitational radiation field can be described as a quantized monochromatic gravitational wave containing $N$ gravitons, then the corresponding gravitoelectric tensor $\epsilon_{ij}$ can be written as \cite{yongs2020prd} 
\beq\label{epsilon ij}
\epsilon_{ij}=-\sqrt{\frac{\hbar G\omega_R^3 \rho_N}{c^2(2\pi)^2 N}} \left[b(\vec k)e_{ij}^{(\varepsilon)} e^{i(\vec k\cdot \vec x-\omega_R t)}+H.c.\right],
\eeq
where $\rho_N$ is the number density of gravitons, $b(\vec k)$ the annihilation operator of the  quantized gravitational wave,  $e_{ij}^{(\varepsilon)}$ the polarization tensors, 
and $\omega_{R}=c|\vec k|$.

The initial state $|\phi\rangle$ of the whole system  is taken as
\beq\label{phi}
|\phi\rangle=|e_0^A\rangle |e_0^B\rangle |0\rangle |N\rangle,
\eeq
where $|0\rangle$ denotes the vacuum state of the fluctuating gravitational field, and $|N\rangle$ is the number state of the external gravitational radiation field. Denote the initial energy of the whole system as $E_{\phi}$.
Then, the  leading-order field-induced interaction energy considered here can be obtained through a direct fourth-order perturbation calculation, i.e.,
\beq
\Delta E_{AB}=\sum_{I_1,I_2,I_3}\frac{\langle \phi|H_I|I_3\rangle\langle I_3|H_I|I_2\rangle\langle I_2|H_I|I_1\rangle\langle I_1|H_I|\phi\rangle}{(E_\phi-E_{I_3})(E_\phi-E_{I_2})(E_\phi-E_{I_1})},
\eeq
which contains  $96$ time-ordered diagrams corresponding to two different kinds of physical processes.
The first kind of processes  is the scattering of  a real graviton  by the two  object centers and an  exchange of a single virtual graviton between them. For a  typical time-ordered diagram, see Fig. 1 in Ref \cite{yongs2020prd}. The  field-induced interaction  associated with this kind of processes is  exactly what we have investigated in Ref. \cite{yongs2020prd}, so we do not repeat it here.
On the other hand, when the objects have permanent quadrupole moments, there is an additional  field-induced interobject interaction, which occurs through the coupling between the permanent quadrupole in one object and the external field-induced quadrupole in the other.
That is, a real graviton is now scattered by the same object center with coupling between the two objects occurring via the exchange of a single virtual graviton. There are $48$ time-ordered diagrams of  this kind,  two typical ones of which are shown in Fig.~\ref{T}.  Summing up all the $48$ possible intermediate processes (see Table \ref{24I} in Appendix \ref{Appendix1}), the additional field-induced interobject interaction energy between two polar objects can then be expressed as 
\begin{figure}[htbp]
  \centering
  \includegraphics[width=0.6\textwidth]{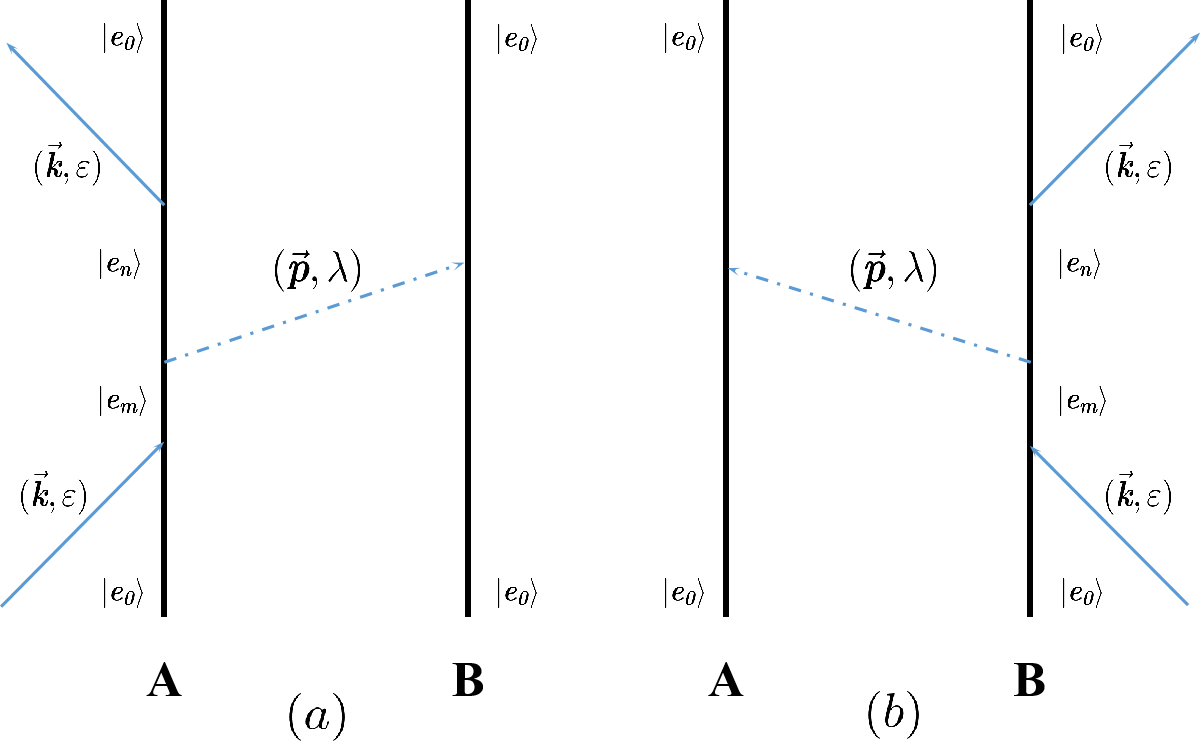}\\
  \caption{Two typical time-ordered diagrams (each with 23 further permutations) for the calculation of the interobject interaction in the existence of an external  gravitational radiation field. The blue solid line represents a real graviton, while the dotted one represents a virtual graviton. }\label{T}
\end{figure}
\bea\label{Eab}
\nonumber \Delta E_{AB}&=&-\frac{\hbar G \omega_R^3 \rho_N}{32\pi^2 c^2} e_{hl}^{(\varepsilon)}e_{ab}^{(\varepsilon)}\hat Q^{B00}_{mn} \sum_{s,t}\Bigg\{\frac{\hat Q^{\xi 0t}_{ij} \hat Q^{\xi ts}_{hl} \hat Q^{\xi s0}_{ab}}{E_{t0}(E_{s0}-\hbar \omega_R)}+\frac{\hat Q^{\xi 0t}_{hl} \hat Q^{\xi ts}_{ij} \hat Q^{\xi s0}_{ab}}{(E_{t0}-\hbar \omega_R)(E_{s0}-\hbar \omega_R)}\\
\nonumber &&+\frac{\hat Q^{\xi 0t}_{hl} \hat Q^{\xi ts}_{ab} \hat Q^{\xi s0}_{ij}}{(E_{t0}-\hbar \omega_R)E_{s0}}
+\frac{N+1}{N}\Bigg[\frac{\hat Q^{A0t}_{ij} \hat Q^{Ats}_{ab} \hat Q^{As0}_{hl}}{E_{t0}(E_{s0}+\hbar \omega_R)}
+\frac{\hat Q^{A0t}_{ab} \hat Q^{Ats}_{ij} \hat Q^{As0}_{hl}}{(E_{t0}+\hbar \omega_R)(E_{s0}+\hbar \omega_R)}\\
&&+\frac{\hat Q^{A0t}_{ab} \hat Q^{Ats}_{hl} \hat Q^{As0}_{ij}}{(E_{t0}+\hbar \omega_R)E_{s0}}\Bigg] \Bigg\} V_{ijmn}(\vec r)
+A\rightleftharpoons B \ terms,
\eea
where $\hat Q^{\xi00}_{ij}=\langle e^{\xi}_0|Q^{\xi}_{ij}|e^{\xi}_0\rangle$ is the permanent quadrupole moment of object $\xi$, 
$\hat Q^{\xi t s}_{ij}=\langle e^{\xi}_t|Q^{\xi}_{ij}|e^{\xi}_s\rangle$ the quadrupole transition moments, and $E_{s0}=E_s-E_0$ the energy level spacing. Here $V_{ijmn}(\vec r)$ is a tensor describes the quadrupole-quadrupole interaction, which is
\beq \label{V_ijmn}
V_{ijmn}(\vec r)=\int d^3\vec p \sum_{\lambda}e^{(\lambda)}_{ij}e^{(\lambda)}_{mn} \frac{G\omega^2 }{4\pi^2 c^2}e^{i\vec p\cdot\vec r},
\eeq
where $\vec r= \vec x_A - \vec x_B$. In the transverse traceless gauge, the summation of polarization tensors gives~\cite{yu1999}
\beq\label{polar-sum}
\sum_{\lambda}e^{(\lambda)}_{ij}e^{(\lambda)}_{mn}=\delta_{i m}\delta_{j n}+\delta_{i n}\delta_{j m}-\delta_{ij}\delta_{m n} -\frac{1}{p^2}H_{ijmn}+\frac{1}{p^4}P_{ijmn},
\eeq
where
\beq
H_{ijmn}=\partial_i\partial_j\delta_{mn}+\partial_m\partial_n\delta_{ij}-\partial_i\partial_m\delta_{jn} -\partial_i\partial_n\delta_{jm}-\partial_j\partial_m\delta_{in}-\partial_j\partial_n\delta_{im},\ P_{ijmn}=\partial_i\partial_j\partial_m\partial_n.
\eeq
Substituting Eq. (\ref{polar-sum}) into Eq. (\ref{V_ijmn}) and performing the integral, it is easy to obtain
\bea\label{eq11}
\nonumber V_{ijmn}(\vec r)=\frac{G}{2 r^5}\Big[&&3(\delta_{i m}\delta_{j n}+\delta_{i n}\delta_{j m}+\delta_{ij}\delta_{m n})-15(\hat r_i \hat r_j \delta_{m n}+\hat r_i \hat r_m \delta_{j n}+\hat r_i \hat r_n \delta_{j m}\\
&&+\hat r_j \hat r_m \delta_{i n}+\hat r_j \hat r_n \delta_{i m}+\hat r_m \hat r_n \delta_{ij})+105\hat r_i \hat r_j \hat r_m \hat r_n \Big],
\eea
where $\hat r_i$ is the $i$th component of the unit vector $\vec r/r$. This shows that,
the additional field-induced interobject interaction between two polar objects in the presence of a weak external gravitational radiation field behaves as $r^{-5}$ in all distance regimes. Notice that, the above result is nonretarded, which is in contrast to the case considered in Ref. \cite{yongs2020prd}. 
This is understandable since, in our case, a real graviton is scattered by the same object center, the process of which is unrelated to the interobject distance, 
while in the case of Ref. \cite{yongs2020prd}, a real graviton is scattered by both the two object centers, the process of which is related to the interobject distance and may lead to a change of the behavior of the interobject interaction  due to the finite speed of graviton propagation. Moreover, from Eq. (\ref{Eab}), we find that such an interaction is related to the number density of gravitons, the frequency and polarization of the external gravitational radiation field, as well as the permanent quadrupoles of the two objects. Note here that, the above calculations are applicable only when the frequency of the external gravitational radiation field is in the off-resonant range.

Now, let us discuss the case in the limit of a large graviton number,  i.e., $N\gg1$, which may correspond to the case of classical gravitational waves. In this regard, the interaction energy shift Eq. (\ref{Eab}) can be expressed as
\beq\label{E_beta}
\Delta E_{AB}\simeq-\frac{I_R}{16}e_{hl}^{(\varepsilon)}e_{ab}^{(\varepsilon)}(\beta^A_{ijhlab}\hat Q^{B00}_{mn}+\beta^B_{ijhlab}\hat Q^{A00}_{mn})V_{ijmn}(\vec r),
\eeq
where $I_R$ is the intensity of the external gravitational radiation field, which is defined as \cite{yongs2020prd}
\beq
I_R=\langle N|\epsilon^2_{ij}|N\rangle=\frac{\hbar G \omega_R^3 \rho_N}{4N\pi^2 c^2}(2N+1)\simeq\frac{\hbar G \omega_R^3 \rho_N}{2\pi^2 c^2},
\eeq
and $\beta^\xi_{ijhlab}$ is the hyperpolarizability of object $\xi$, which is defined in analogy to the electromagnetic case \cite{Andrews2000,Andrews2005,Salam2007} as
\bea\label{beta}
\nonumber\beta_{ijhlab}^\xi(\omega_R)&=& \sum_{s,t}\Bigg[\frac{\hat Q^{\xi 0t}_{ij} \hat Q^{\xi ts}_{hl} \hat Q^{\xi s0}_{ab}}{E_{t0}(E_{s0}-\hbar \omega_R)}+\frac{\hat Q^{\xi 0t}_{hl} \hat Q^{\xi ts}_{ij} \hat Q^{\xi s0}_{ab}}{(E_{t0}-\hbar \omega_R)(E_{s0}-\hbar \omega_R)}\\
\nonumber &&+\frac{\hat Q^{\xi 0t}_{hl} \hat Q^{\xi ts}_{ab} \hat Q^{\xi s0}_{ij}}{(E_{t0}-\hbar \omega_R)E_{s0}}
+\frac{\hat Q^{\xi 0t}_{ab} \hat Q^{\xi ts}_{ij} \hat Q^{\xi s0}_{hl}}{(E_{t0}+\hbar \omega_R)(E_{s0}+\hbar \omega_R)}\\
&& +\frac{\hat Q^{\xi 0t}_{ij} \hat Q^{\xi ts}_{ab} \hat Q^{\xi s0}_{hl}}{E_{t0}(E_{s0}+\hbar \omega_R)}
+\frac{\hat Q^{\xi 0t}_{ab} \hat Q^{\xi ts}_{hl} \hat Q^{\xi s0}_{ij}}{(E_{t0}+\hbar \omega_R)E_{s0}}\Bigg],
\eea
Thus, in the  large graviton number limit, the interobject interaction is now determined by the permanent quadrupole moments and the hyperpolarizabilities of the two objects, as well as the intensity and polarization of the external gravitational wave.
In this aspect, such an interaction can be understood intuitively from the point of induced quadrupole moments. That is, the externally applied gravitational wave induces a quadrupole moment which is quadratic in the field in one object, which then interacts with the permanent quadrupole moment in the other object via the static quadrupole-quadrupole interaction tensor (i.e., $V_{ijmn}$), and an interaction energy is thus obtained.

A few comments are now in order. First, we compare the field-induced interobject  interaction  between two  polar objects $\Delta E_{polar}$ with that between  nonpolar ones $\Delta E_{non}$. 
As discussed before,  for the polar case, there exists an additional term which is related to the hyperpolarizability $\beta_{ijhlab}$ and here labeled as $\Delta E_{\beta}$, due to the scattering of a real graviton by the same object center and an exchange of a single virtual graviton, i.e., $\Delta E_{polar}=\Delta E_{non}+\Delta E_{\beta}$.
From Ref. \cite{yongs2020prd}, we know that the interaction between two nonpolar objects can be either attractive or repulsive depending on the propagation direction and polarization of the external gravitational field. However, as our main concern here is the relative magnitude of the extrema of the interaction potentials in the polar and nonpolar cases, we rewrite the extremum of the interaction potential $\Delta E_{non}$ as 
\begin{equation}\label{E_alpha non}
\Delta E_{non}\sim\left\{
\begin{aligned}
&-\frac{\hbar c G^2\rho_N}{\lambda_R^3 r^5}\alpha^2(\omega_R), \qquad r\ll\lambda_R,\\
&-\frac{\hbar c G^2\rho_N}{\lambda_R^7 r}\alpha^2(\omega_R),\qquad r\gg\lambda_R,
\end{aligned}\right.
\end{equation}
where $\lambda_R$ is the wavelength of the external gravitational field, and $\alpha(\omega_R)$ is the isotropic gravitational quadrupole polarizability of the object, which takes the form
\beq\label{alpha ck}
\alpha(\omega_R)=\sum_{s}\frac{E_{s0} \hat Q^{0s}\hat Q^{s0}}{E_{s0}^2-(\hbar \omega_R)^2} =\sum_{s}\frac{\alpha_s(0)}{1-(\hbar \omega_R/E_{s0})^2},
\eeq
with $\alpha(0)=\sum_s\alpha_s(0)$ being the static polarizability (i.e., the case when $k=0$).
From Eq. (\ref{E_beta}), $\Delta E_{\beta}$    can be approximately written as
\beq\label{E_beta2}
\Delta E_{\beta}\sim-\frac{\hbar c G^2\rho_N}{\lambda_R^3 r^5}\beta(\omega_R)\hat Q^{00},
\eeq
where $\hat Q^{00}$ is the permanent quadrupole moment, and $\beta(\omega_R)$ denotes the leading (second) term of the gravitational quadrupole hyperpolarizability given in Eq. (\ref{beta}), which can be  approximately expressed as
\beq
\beta(\omega_R)\sim\sum_{s,t}\frac{\beta_{s,t}(0)}{(1-\hbar \omega_R/E_{s0})(1-\hbar \omega_R/E_{t0})},
\eeq
with $\beta(0)\sim\sum_{s,t}\beta_{s,t}(0)$ being the static hyperpolarizability. 
In the near regime where the interobject distance is much smaller than the wavelength of the  external gravitational radiation field, the ratio of $\Delta E_{polar}$ to  $\Delta E_{non}$ can be approximately obtained as
\beq
\frac{\Delta E_{polar}}{\Delta E_{non}} 
\sim1+\frac{\beta(0)\hat Q^{00}} {\alpha^2(0)}.
\eeq
Obviously, when the permanent quadrupole $\hat Q^{00}$ is much larger than the  characteristic quadrupole described by the ratio of two polarizabilities, i.e,  $\frac{\alpha^2(0)}{\beta(0)}$, the external gravitational field-induced interaction between two polar objects is significantly different from that between the nonpolar ones.
In the far regime, since $\Delta E_{non}$ behaves as $r^{-1}$ while $\Delta E_{\beta}$ behaves as $r^{-5}$, the difference between the field-induced  interactions in the two cases is usually negligible.
However, at some special positions where  $\Delta E_{non}$ is zero, the interaction between nonpolar objects  vanishes but that between polar ones are not.

Second, let us note here that although we have modeled the external gravitational radiation as a quantized monochromatic gravitational wave with a certain wave vector and polarization for simplicity, it is possible to generalize to more realistic situations. As an example, in the following, we consider  the case of a stochastic background of gravitational radiation.
Summing all the propagation directions, polarizations and frequencies of the gravitational waves, the  interaction potential between two objects with permanent quadrupole moments induced by a stochastic background of gravitational waves can be formally expressed as
\bea
\nonumber\Delta E_{stoch}&=&\int f(\omega_R)d \omega_R \int d\Omega_{\vec k} \sum_{\varepsilon} \Delta E_{polar} \\
\nonumber&=&-\frac{1}{8} \int I_R(\omega_R) f(\omega_R) d \omega_R
\int d\Omega_{\vec k} \sum_{\varepsilon} \Bigg[\alpha^{A}_{ijab}\alpha^{B}_{hlmn}e^{(\varepsilon)}_{ab}e^{(\varepsilon)}_{mn} \cos{(\vec k \cdot \vec r)} V_{ijhl}(\omega_R,\vec r)\\
&&+\frac{1}{2} e_{hl}^{(\varepsilon)}e_{ab}^{(\varepsilon)} (\beta^A_{ijhlab}\hat Q^{B00}_{mn}+\beta^B_{ijhlab}\hat Q^{A00}_{mn})V_{ijmn}(\vec r)\Bigg],
\eea
where $f(\omega_R)$ is the frequency distribution function of the stochastic gravitational waves, $\int d \Omega_{\vec k}$ denotes the integral over solid angle, $V_{ijkl}(\omega_R,\vec r)$ is the dynamic quadrupole-quadrupole interaction tensor given in Ref. \cite{yongs2020prd} (See Eq. (18) in Ref. \cite{yongs2020prd}), and $V_{ijkl}(\vec r)$ is the static quadrupole-quadrupole interaction tensor given in Eq. \eqref{eq11}. That is, one may treat the spectrum of the stochastic gravitational waves as a frequency-dependent distribution function and the induced interobject interaction potential can then be obtained via an integral over the frequencies as well as the summations of the propagation directions and polarizations.

\section{Summary }
\label{sec_disc}

Within the framework of linearized quantum gravity, the induced quantum gravitational interaction between two ground-state objects with permanent quadrupole moments, which are coupled with the fluctuating gravitational fields in vacuum and subjected to a weak external gravitational radiation field, is investigated based on the leading-order perturbation theory. Compared with the field-induced interaction between nonpolar objects,  there exists an additional term in the leading-order field-induced interobject interaction between two polar objects, which behaves as $r^{-5}$ in all distance regimes and arises from  the scattering of a real graviton  by the same object center with coupling between the two objects occurring via the exchange of a single virtual graviton. Such an additional interaction is relevant to the number density of gravitons, the polarization and frequency of the external gravitational field, as well as the permanent quadrupoles of the objects.
In the limit of a large number of gravitons, the interaction can be viewed as the interaction between the induced gravitational quadrupole in one object with the permanent quadrupole moment in the other object, which can be characterized by the permanent quadrupole moments and the hyperpolarizabilities of the two objects, as well as the intensity and polarization of the external gravitational wave.
Due to the existence of such an additional term, the field-induced quantum gravitational interaction between two polar objects can be significantly different from that between nonpolar ones when the interobject distance is much smaller than the wavelength of the  external gravitational radiation field.
Although we model the external gravitational radiation as a quantized monochromatic gravitational wave with a certain wave vector and polarization for simplicity, it is possible to generalize to more realistic situations, such as the case of a stochastic background of gravitational radiation.

\begin{acknowledgments}

This work was supported in part by the NSFC under Grants No. 11690034, No. 11805063 and No. 12075084, and the Hunan Provincial Natural Science Foundation of China under Grant No. 2020JJ3026.

\end{acknowledgments}


\appendix
\section{Intermediate processes}\label{Appendix1}
The intermediate states and their associated energy denominators of Eq. (\ref{Eab}) are followed.

\begin{table}[H]
  \centering
  \scalebox{0.75}{
\begin{tabular}{cllll}
\hline
Case  &\hspace{1ex} $|\text{I}\rangle$ &\hspace{2ex}$|\text{II}\rangle$ &\hspace{2ex}$|\text{III}\rangle$ &\hspace{2ex} Denominator\\
\hline
(1) &\hspace{1ex} $|e_m^A\rangle |e_0^B\rangle |0\rangle |N-1\rangle$
      & \hspace{2ex}$|e_s^A\rangle |e_0^B\rangle |0\rangle |N\rangle$
      & \hspace{2ex}$|e_0^A\rangle |e_0^B\rangle |1\rangle |N\rangle$
      & \hspace{2ex}$D_{1}=(E^A_{m0}-\hbar \omega_R) E^A_{s0} \hbar\omega$  \\
(2) &\hspace{1ex} $|e_m^A\rangle |e_0^B\rangle |0\rangle |N-1\rangle$
      & \hspace{2ex}$|e_s^A\rangle |e_0^B\rangle |0\rangle |N\rangle$
      & \hspace{2ex}$|e_s^A\rangle |e_0^B\rangle |1\rangle |N\rangle$
      & \hspace{2ex}$D_{2}=(E^A_{m0}-\hbar \omega_R) E^A_{s0} (E^A_{s0}+\hbar\omega)$  \\
(3) &\hspace{1ex} $|e_m^A\rangle |e_0^B\rangle |0\rangle |N-1\rangle$
      & \hspace{2ex}$|e_m^A\rangle |e_0^B\rangle |1\rangle |N-1\rangle$
      & \hspace{2ex}$|e_s^A\rangle |e_0^B\rangle |1\rangle |N\rangle$
      & \hspace{2ex}$D_{3}=(E^A_{m0}-\hbar \omega_R) (E^A_{m0}-\hbar \omega_R+\hbar\omega) (E^A_{s0}+\hbar\omega)$  \\
(4) &\hspace{1ex} $|e_0^A\rangle |e_0^B\rangle |1\rangle |N\rangle$
      & \hspace{2ex}$|e_m^A\rangle |e_0^B\rangle |1\rangle |N-1\rangle$
      & \hspace{2ex}$|e_s^A\rangle |e_0^B\rangle |1\rangle |N\rangle$
      & \hspace{2ex}$D_{4}=\hbar\omega (E^A_{m0}-\hbar \omega_R+\hbar\omega) (E^A_{s0}+\hbar\omega)$  \\
(5) &\hspace{1ex} $|e_m^A\rangle |e_0^B\rangle |0\rangle |N-1\rangle$
      & \hspace{2ex}$|e_s^A\rangle |e_0^B\rangle |1\rangle |N-1\rangle$
      & \hspace{2ex}$|e_0^A\rangle |e_0^B\rangle |1\rangle |N\rangle$
      & \hspace{2ex}$D_{5}=(E^A_{m0}-\hbar \omega_R) (E^A_{s0}-\hbar \omega_R+\hbar\omega) \hbar\omega$  \\
(6) &\hspace{1ex} $|e_m^A\rangle |e_0^B\rangle |0\rangle |N-1\rangle$
      & \hspace{2ex}$|e_s^A\rangle |e_0^B\rangle |1\rangle |N-1\rangle$
      & \hspace{2ex}$|e_s^A\rangle |e_0^B\rangle |0\rangle |N-1\rangle$
      & \hspace{2ex}$D_{6}=(E^A_{m0}-\hbar \omega_R) (E^A_{s0}-\hbar \omega_R+\hbar\omega) (E^A_{s0}-\hbar \omega_R)$  \\
(7) &\hspace{1ex} $|e_m^A\rangle |e_0^B\rangle |0\rangle |N-1\rangle$
      & \hspace{2ex}$|e_m^A\rangle |e_0^B\rangle |1\rangle |N-1\rangle$
      & \hspace{2ex}$|e_s^A\rangle |e_0^B\rangle |0\rangle |N-1\rangle$
      & \hspace{2ex}$D_{7}=(E^A_{m0}-\hbar \omega_R) (E^A_{m0}-\hbar \omega_R+\hbar\omega) (E^A_{s0}-\hbar \omega_R)$  \\
(8) &\hspace{1ex} $|e_0^A\rangle |e_0^B\rangle |1\rangle |N\rangle$
      & \hspace{2ex}$|e_m^A\rangle |e_0^B\rangle |1\rangle |N-1\rangle$
      & \hspace{2ex}$|e_s^A\rangle |e_0^B\rangle |0\rangle |N-1\rangle$
      & \hspace{2ex}$D_{8}=\hbar\omega (E^A_{m0}-\hbar \omega_R+\hbar\omega) (E^A_{s0}-\hbar \omega_R)$  \\
(9) &\hspace{1ex} $|e_0^A\rangle |e_0^B\rangle |1\rangle |N\rangle$
      & \hspace{2ex}$|e_m^A\rangle |e_0^B\rangle |0\rangle |N\rangle$
      & \hspace{2ex}$|e_s^A\rangle |e_0^B\rangle |0\rangle |N-1\rangle$
      & \hspace{2ex}$D_{9}=\hbar\omega E^A_{m0} (E^A_{s0}-\hbar \omega_R)$  \\
(10)&\hspace{1ex} $|e_m^A\rangle |e_0^B\rangle |1\rangle |N\rangle$
      & \hspace{2ex}$|e_m^A\rangle |e_0^B\rangle |0\rangle |N\rangle$
      & \hspace{2ex}$|e_s^A\rangle |e_0^B\rangle |0\rangle |N-1\rangle$
      & \hspace{2ex}$D_{10}=(E^A_{m0}+\hbar\omega) E^A_{m0} (E^A_{s0}-\hbar \omega_R)$  \\
(11)&\hspace{1ex} $|e_m^A\rangle |e_0^B\rangle |1\rangle |N\rangle$
      & \hspace{2ex}$|e_s^A\rangle |e_0^B\rangle |1\rangle |N-1\rangle$
      & \hspace{2ex}$|e_s^A\rangle |e_0^B\rangle |0\rangle |N-1\rangle$
      & \hspace{2ex}$D_{11}=(E^A_{m0}+\hbar\omega) (E^A_{s0}-\hbar \omega_R+\hbar\omega) (E^A_{s0}-\hbar \omega_R)$  \\
(12)&\hspace{1ex} $|e_m^A\rangle |e_0^B\rangle |1\rangle |N\rangle$
      & \hspace{2ex}$|e_s^A\rangle |e_0^B\rangle |1\rangle |N-1\rangle$
      & \hspace{2ex}$|e_0^A\rangle |e_0^B\rangle |1\rangle |N\rangle$
      & \hspace{2ex}$D_{12}=(E^A_{m0}+\hbar\omega) (E^A_{s0}-\hbar \omega_R+\hbar\omega) \hbar\omega$  \\
(13)&\hspace{1ex} $|e_m^A\rangle |e_0^B\rangle |0\rangle |N+1\rangle$
      & \hspace{2ex}$|e_s^A\rangle |e_0^B\rangle |0\rangle |N\rangle$
      & \hspace{2ex}$|e_0^A\rangle |e_0^B\rangle |1\rangle |N\rangle$
      & \hspace{2ex}$D_{13}=(E^A_{m0}+\hbar \omega_R) E^A_{s0} \hbar\omega$  \\
(14)&\hspace{1ex} $|e_m^A\rangle |e_0^B\rangle |0\rangle |N+1\rangle$
      & \hspace{2ex}$|e_s^A\rangle |e_0^B\rangle |0\rangle |N\rangle$
      & \hspace{2ex}$|e_s^A\rangle |e_0^B\rangle |1\rangle |N\rangle$
      & \hspace{2ex}$D_{14}=(E^A_{m0}+\hbar \omega_R) E^A_{s0} (E^A_{s0}+\hbar\omega)$  \\
(15)&\hspace{1ex} $|e_m^A\rangle |e_0^B\rangle |0\rangle |N+1\rangle$
      & \hspace{2ex}$|e_m^A\rangle |e_0^B\rangle |1\rangle |N+1\rangle$
      & \hspace{2ex}$|e_s^A\rangle |e_0^B\rangle |1\rangle |N\rangle$
      & \hspace{2ex}$D_{15}=(E^A_{m0}+\hbar \omega_R) (E^A_{m0}+\hbar \omega_R+\hbar\omega) (E^A_{s0}+\hbar\omega)$  \\
(16)&\hspace{1ex} $|e_0^A\rangle |e_0^B\rangle |1\rangle |N\rangle$
      & \hspace{2ex}$|e_m^A\rangle |e_0^B\rangle |1\rangle |N+1\rangle$
      & \hspace{2ex}$|e_s^A\rangle |e_0^B\rangle |1\rangle |N\rangle$
      & \hspace{2ex}$D_{16}=\hbar\omega (E^A_{m0}+\hbar \omega_R+\hbar\omega) (E^A_{s0}+\hbar\omega)$  \\
(17)&\hspace{1ex} $|e_m^A\rangle |e_0^B\rangle |0\rangle |N+1\rangle$
      & \hspace{2ex}$|e_s^A\rangle |e_0^B\rangle |1\rangle |N+1\rangle$
      & \hspace{2ex}$|e_0^A\rangle |e_0^B\rangle |1\rangle |N\rangle$
      & \hspace{2ex}$D_{17}=(E^A_{m0}+\hbar \omega_R) (E^A_{s0}+\hbar \omega_R+\hbar\omega) \hbar\omega$  \\
(18)&\hspace{1ex} $|e_m^A\rangle |e_0^B\rangle |0\rangle |N+1\rangle$
      & \hspace{2ex}$|e_s^A\rangle |e_0^B\rangle |1\rangle |N+1\rangle$
      & \hspace{2ex}$|e_s^A\rangle |e_0^B\rangle |0\rangle |N+1\rangle$
      & \hspace{2ex}$D_{18}=(E^A_{m0}+\hbar \omega_R) (E^A_{s0}+\hbar \omega_R+\hbar\omega) (E^A_{s0}+\hbar \omega_R)$  \\
(19)&\hspace{1ex} $|e_m^A\rangle |e_0^B\rangle |0\rangle |N+1\rangle$
      & \hspace{2ex}$|e_m^A\rangle |e_0^B\rangle |1\rangle |N+1\rangle$
      & \hspace{2ex}$|e_s^A\rangle |e_0^B\rangle |0\rangle |N+1\rangle$
      & \hspace{2ex}$D_{19}=(E^A_{m0}+\hbar \omega_R) (E^A_{m0}+\hbar \omega_R+\hbar\omega) (E^A_{s0}+\hbar \omega_R)$  \\
(20)&\hspace{1ex} $|e_0^A\rangle |e_0^B\rangle |1\rangle |N\rangle$
      & \hspace{2ex}$|e_m^A\rangle |e_0^B\rangle |1\rangle |N+1\rangle$
      & \hspace{2ex}$|e_s^A\rangle |e_0^B\rangle |0\rangle |N+1\rangle$
      & \hspace{2ex}$D_{20}=\hbar\omega (E^A_{m0}+\hbar \omega_R+\hbar\omega) (E^A_{s0}+\hbar \omega_R)$  \\
(21)&\hspace{1ex} $|e_0^A\rangle |e_0^B\rangle |1\rangle |N\rangle$
      & \hspace{2ex}$|e_m^A\rangle |e_0^B\rangle |0\rangle |N\rangle$
      & \hspace{2ex}$|e_s^A\rangle |e_0^B\rangle |0\rangle |N+1\rangle$
      & \hspace{2ex}$D_{21}=\hbar\omega E^A_{m0} (E^A_{s0}+\hbar \omega_R)$  \\
(22)&\hspace{1ex} $|e_m^A\rangle |e_0^B\rangle |1\rangle |N\rangle$
      & \hspace{2ex}$|e_m^A\rangle |e_0^B\rangle |0\rangle |N\rangle$
      & \hspace{2ex}$|e_s^A\rangle |e_0^B\rangle |0\rangle |N+1\rangle$
      & \hspace{2ex}$D_{22}=(E^A_{m0}+\hbar\omega) E^A_{m0} (E^A_{s0}+\hbar \omega_R)$  \\
(23)&\hspace{1ex} $|e_m^A\rangle |e_0^B\rangle |1\rangle |N\rangle$
      & \hspace{2ex}$|e_s^A\rangle |e_0^B\rangle |1\rangle |N+1\rangle$
      & \hspace{2ex}$|e_s^A\rangle |e_0^B\rangle |0\rangle |N+1\rangle$
      & \hspace{2ex}$D_{23}=(E^A_{m0}+\hbar\omega) (E^A_{s0}+\hbar \omega_R+\hbar\omega) (E^A_{s0}+\hbar \omega_R)$  \\
(24)&\hspace{1ex} $|e_m^A\rangle |e_0^B\rangle |1\rangle |N\rangle$
      & \hspace{2ex}$|e_s^A\rangle |e_0^B\rangle |1\rangle |N+1\rangle$
      & \hspace{2ex}$|e_0^A\rangle |e_0^B\rangle |1\rangle |N\rangle$
      & \hspace{2ex}$D_{24}=(E^A_{m0}+\hbar\omega) (E^A_{s0}+\hbar \omega_R+\hbar\omega) \hbar\omega$  \\
\hline
\end{tabular}}
  \caption{Intermediate states and the corresponding energy denominators for the case when a real graviton is scattered by object A. Similar processes for object B can be obtained by exchanging the label A and B.}\label{24I}
\end{table}


\begin{thebibliography}{99}
\bibitem{Einstein1916} A. Einstein, \textit{Approximate integration of field equations of gravitation}, Sitzungsber. K. Preuss. Akad. Wiss. {\bf 1}, 688 (1916).

\bibitem{Donoghue1994prl} J. F. Donoghue, \textit{Leading quantum correction to the Newtonian potential},  \href{https://doi.org/10.1103/PhysRevLett.72.2996}{Phys. Rev. Lett. {\bf 72}, 2996 (1994)}.
\bibitem{Donoghue1994prd} J. F. Donoghue, \textit{General relativity as an effective field theory: The leading quantum corrections}, \href{https://doi.org/10.1103/PhysRevD.50.3874}{Phys. Rev. D {\bf 50}, 3874 (1994)}.
\bibitem{Hamber1995} H. W. Hamber and S. Liu, \textit{On the quantum corrections to the newtonian potential}, \href{https://doi.org/10.1016/0370-2693(95)00790-R}{Phys. Lett. B {\bf 357}, 51 (1995)}.
\bibitem{Kirilin2002} I. B. Khriplovich and G. G. Kirilin, \textit{Quantum power correction to the Newton law}, Zh. Eksp. Teor. Fiz. {\bf 95}, 1139 (2002) [\href{https://doi.org/10.1134/1.1537290}{J. Exp. Theor. Phys. {\bf 95}, 981 (2002)}].
\bibitem{Holstein2003} N. E. J. Bjerrum-Bohr, J. F. Donoghue, and B. R. Holstein, \textit{Quantum gravitational corrections to the nonrelativistic scattering potential of two masses}, \href{https://doi.org/10.1103/PhysRevD.67.084033}{Phys. Rev. D {\bf 67}, 084033 (2003)}.
\bibitem{Holstein2005} N. E. J. Bjerrum-Bohr, J. F. Donoghue, and B. R. Holstein, \textit{Erratum: Quantum gravitational corrections to the nonrelativistic scattering potential of two masses [Phys. Rev. D 67, 084033 (2003)]}, \href{https://doi.org/10.1103/PhysRevD.71.069903}{Phys. Rev. D {\bf 71}, 069903(E) (2005)}.



\bibitem{Ford1995} L. H. Ford, \textit{Gravitons and light cone fluctuations}, \href{https://doi.org/10.1103/PhysRevD.51.1692}{Phys. Rev. D {\bf 51}, 1692 (1995)}.
\bibitem{Ford1996} L. H. Ford and N. F. Svaiter, \textit{Gravitons and light cone fluctuations. II. Correlation functions}, \href{https://doi.org/10.1103/PhysRevD.54.2640}{Phys. Rev. D {\bf 54}, 2640 (1996)}.
\bibitem{yu1999} H. Yu and L. H. Ford, \textit{Light-cone fluctuations in flat spacetimes with nontrivial topology}, \href{https://doi.org/10.1103/PhysRevD.60.084023}{Phys. Rev. D {\bf 60}, 084023 (1999)}.
\bibitem{yu2000} H. Yu and L. H. Ford, \textit{Lightcone fluctuations in quantum gravity and extra dimensions}, \href{https://doi.org/10.1016/S0370-2693(00)01287-9}{Phys. Lett. B {\bf 496}, 107 (2000)}.
\bibitem{yu2009} H. Yu, N. F. Svaiter, and L. H. Ford, \textit{Quantum light-cone fluctuations in compactified spacetimes}, \href{https://doi.org/10.1103/PhysRevD.80.124019}{Phys. Rev. D {\bf 80}, 124019 (2009)}.



\bibitem{Ford2016} L. H. Ford, M. P. Hertzberg, and J. Karouby, \textit{Quantum Gravitational Force Between Polarizable Objects}, \href{https://doi.org/10.1103/PhysRevLett.116.151301}{ Phys. Rev. Lett. {\bf 116}, 151301 (2016)}.
\bibitem{Wu2016} P. Wu, J. Hu, and H. Yu, \textit{Quantum correction to classical gravitational interaction between two polarizable objects}, \href{https://doi.org/10.1016/j.physletb.2016.10.025}{ Phys. Lett. B {\bf 763}, 40 (2016)}.
\bibitem{Wu2017} P. Wu, J. Hu, and H. Yu, \textit{Interaction between two gravitationally polarizable objects induced by thermal bath of gravitons}, \href{https://doi.org/10.1103/PhysRevD.95.104057}{Phys. Rev. D {\bf 95}, 104057 (2017)}.
\bibitem{Holstein2017} B. R. Holstein, \textit{Analytical on-shell calculation of low energy higher order scattering}, \href{https://doi.org/10.1088/0954-3899/44/1/01LT01}{J. Phys. G {\bf 44}, 01LT01 (2017)}.
\bibitem{Hu2017} J. Hu and H. Yu, \textit{Gravitational Casimir-Polder effect}, \href{https://doi.org/10.1016/j.physletb.2017.01.038}{ Phys. Lett. B {\bf 767}, 16 (2017)}.
\bibitem{yu2018} H. Yu, Z. Yang, and P. Wu, \textit{Quantum interaction between two gravitationally polarizable objects in the presence of boundaries}, \href{https://doi.org/10.1103/PhysRevD.97.026008}{Phys. Rev. D {\bf 97}, 026008 (2018)}.




\bibitem{CP} H. B. G. Casimir and D. Polder, \textit{The Influence of Retardation on the London-van der Waals Forces}, \href{https://doi.org/10.1103/PhysRev.73.360}{Phys. Rev. {\bf 73}, 360 (1948)}.
\bibitem{PT} D. P. Craig and T. Thirunamachandran, {\it Molecular Quantum Electrodynamics}  (Dover, Mineola, 1998).
\bibitem{Salam} A. Salam, {\it Molecular Quantum Electrodynamics} (Wiley, Hoboken, NJ, 2010).
\bibitem{Thirunamachandran1980} T. Thirunamachandran, \textit{Intermolecular interactions in the presence of an intense radiation field}, \href{https://doi.org/10.1080/00268978000101561}{Mol. Phys. {\bf 40}, 393 (1980)}.
\bibitem{Milonni1992} P. W. Milonni and M.-L. Shih, \textit{Source theory of the Casimir force}, \href{https://doi.org/10.1103/PhysRevA.45.4241}{Phys. Rev. A {\bf 45}, 4241 (1992)}.
\bibitem{Milonni1996}P. W. Milonni and A. Smith, \textit{van der Waals dispersion forces in electromagnetic fields}, \href{https://doi.org/10.1103/PhysRevA.53.3484}{Phys. Rev. A {\bf 53}, 3484 (1996)}.
\bibitem{Salam2006} A. Salam, \textit{Intermolecular interactions in a radiation field via the method of induced moments}, \href{https://doi.org/10.1103/PhysRevA.73.013406}{Phys. Rev. A {\bf 73}, 013406 (2006)}.
\bibitem{Buhmann2019} T. Haug, S. Y. Buhmann, and R. Bennett, \textit{Casimir-Polder potential in the presence of a Fock state}, \href{https://doi.org/10.1103/PhysRevA.99.012508}{Phys. Rev. A {\bf 99}, 012508 (2019)}.



\bibitem{yongs2020epjc} Y. Hu, J. Hu, H. Yu, and P. Wu, \textit{Resonance interaction between two entangled gravitational polarizable objects}, \href{https://doi.org/10.1140/epjc/s10052-020-8375-y}{Eur. Phys. J. C {\bf 80}, 792 (2020)}.
\bibitem{yongs2020prd} Y. Hu, J. Hu, and H. Yu, \textit{Quantum gravitational interaction between two objects induced by external gravitational radiation fields}, \href{https://doi.org/10.1103/PhysRevD.101.066015}{Phys. Rev. D {\bf 101}, 066015 (2020)}.





\bibitem{Andrews2000} P. Allcock, R. D. Jenkins, and D. L. Andrews, \textit{Laser-assisted resonance-energy transfer}, \href{https://doi.org/10.1103/PhysRevA.61.023812}{Phys. Rev. A {\bf 61}, 023812 (2000)}.
\bibitem{Andrews2005} D. S. Bradshaw and David L. Andrews, \textit{Optically induced forces and torques: Interactions between nanoparticles in a laser beam}, \href{https://doi.org/10.1103/PhysRevA.72.033816}{Phys. Rev. A {\bf 72}, 033816 (2005)}.
\bibitem{Salam2007} A. Salam, \textit{Two alternative derivations of the static contribution to the radiation-induced intermolecular energy shift}, \href{https://doi.org/10.1103/PhysRevA.76.063402}{Phys. Rev. A {\bf 76}, 063402 (2007)}.



\bibitem{Campbell1976} W. B. Campbell and T. A. Morgan, \textit{Maxwell form of the linear theory of gravitation}, \href{https://doi.org/10.1119/1.10195}{Am. J. Phys. {\bf 44}, 356 (1976)}.
\bibitem{Matte1953} A. Matte, \textit{Sur De Nouvelles Solutions Oscillatoires DesEquations De La Gravitation}, \href{https://doi.org/10.4153/CJM-1953-001-3}{Can. J. Math. {\bf 5}, 1 (1953)}.
\bibitem{Campbell1971} W. B. Campbell and T. Morgan, \textit{Debye potentials for the gravitational field}, \href{https://doi.org/10.1016/0031-8914(71)90074-7}{Physica (Utrecht) {\bf 53}, 264 (1971)}.
\bibitem{Szekeres1971} P. Szekeres,  \textit{Linearized gravitation theory in macroscopic media}, \href{https://doi.org/10.1016/0003-4916(71)90117-5}{Ann. Phys. (N.Y.) {\bf 64}, 599 (1971)}.
\bibitem{Maartens1998} R. Maartens and B. A. Bassett, \textit{Gravito-electromagnetism}, \href{https://doi.org/10.1088/0264-9381/15/3/018}{Classical Quantum Gravity {\bf 15}, 705 (1998)}.
\bibitem{Ruggiero2002} M. L. Ruggiero and A. Tartaglia, \textit{Gravitomagnetic effects}, Nuovo Cimento Soc. Ital. Fis. {\bf 117B}, 743 (2002).
\bibitem{Ramos2010} J. Ramos, M. de Montigny, and F. Khanna, \textit{On a Lagrangian formulation of gravitoelectromagnetism}, \href{https://doi.org/10.1007/s10714-010-0990-8}{Gen. Relativ. Gravit. {\bf 42}, 2403 (2010)}.


\end{thebibliography}
\end{document}